\documentclass[twocolumn,showpacs,amsmath,amssymb,prb,superscriptaddress]{revtex4}
\usepackage{bm}
\usepackage[normalem]{ulem}

\usepackage{graphicx}
\usepackage{color}
\usepackage{hhline}
\usepackage{booktabs}
\usepackage{amssymb}

\usepackage{epstopdf}

\begin{document}

\newcommand{\IJMPB}{\textit{Int. J. Mod. Phys. B} }
\newcommand{\PhC}{\textit{Physica C} }
\newcommand{\PhB}{\textit{Physica B} }
\newcommand{\JS}{\textit{J. Supercond. Nov. Magn.} }
\newcommand{\IEEEmw}{\textit{IEEE Trans. Microwave Theory Tech.} }
\newcommand{\IEEEas}{\textit{IEEE Trans. Appl. Supercond.} }
\newcommand{\IEEEim}{\textit{IEEE Trans. Instr. Meas.} }
\newcommand{\PRB}{\textit{Phys. Rev. B} }
\newcommand{\PRL}{{Phys. Rev. Lett.} }
\newcommand{\RMP}{\textit{Rev. Mod. Phys.} }
\newcommand{\IJIMW}{\textit{Int. J. Infrared Millim. Waves} }
\newcommand{\APL}{\textit{Appl. Phys. Lett.} }
\newcommand{\JAP}{{J. Appl. Phys.} }
\newcommand{\JPCM}{{J. Phys.: Condens. Matter} }
\newcommand{\AdP}{\textit{Adv. Phys.} }
\newcommand{\Nat}{\textit{Nature} }
\newcommand{\CM}{\textit{cond-mat/} }
\newcommand{\JpnJAP}{\textit{Jpn. J. Appl. Phys.} }
\newcommand{\PhT}{\textit{Phys. Today} }
\newcommand{\ZETF}{\textit{Zh. Eksperim. i. Teor. Fiz.} }
\newcommand{\JETP}{\textit{Soviet Phys.--JETP} }
\newcommand{\EL}{\textit{Europhys. Lett.} }
\newcommand{\Sci}{\textit{Science} }
\newcommand{\EPJB}{\textit{Eur. Phys. J. B} }
\newcommand{\SUST}{{Supercond. Sci. Technol.} }

\newcommand{\red}[1]{ \textcolor{red}{#1} }

\title{Robustness of the $0-\pi$ transition against compositional and structural ageing in S/F/S heterostructures}
\author{R.~Loria}
\author{C.~Meneghini}
\affiliation{Dipartimento di Scienze, Universit\`a Roma Tre,
Via della Vasca Navale 84, 00146 Roma, Italy}
\author{K.~Torokhtii}
\affiliation{Dipartimento di Ingegneria,
Universit\`a Roma Tre, Via della Vasca Navale 84, 00146 Roma,
Italy}
\author{L.~Tortora}
\affiliation{INFN and Dipartimento di Matematica e Fisica,
Universit\`a  Roma Tre, Via della Vasca Navale 84, 00146 Roma,
Italy}
\author{N.~Pompeo}
\affiliation{Dipartimento di Ingegneria,
Universit\`a Roma Tre, Via della Vasca Navale 84, 00146 Roma,
Italy}
\author{C.~Cirillo}
\author{C.~Attanasio}
\affiliation{CNR-SPIN and Dipartimento di Fisica ``E. Caianiello'',
Universit\`a  di Salerno, 84084 Fisciano (SA),
Italy}
\author{E.~Silva}
\email{enrico.silva@uniroma3.it}
\affiliation{Dipartimento di Ingegneria,
Universit\`a Roma Tre, Via della Vasca Navale 84, 00146 Roma,
Italy}

\date{\today}

\begin{abstract}
We have studied the temperature induced $0 -\pi $ thermodynamic transition in Nb/PdNi/Nb Superconductor/Ferromagnetic/Superconductor (SFS) heterostructures by microwave measurements of the superfluid density. We have observed a shift in the transition temperature with the ageing of the heterostructures, suggesting that structural and/or chemical changes took place. Motivated by the electrodynamics findings, we have extensively studied the local structural properties of the samples by means of X-ray Absorption Spectroscopy (XAS) technique, and the compositional profile by Time--of--Flight Secondary Ion Mass Spectrometry (ToF--SIMS). We found that the samples have indeed changed their properties, in particular for what concerns the interfaces and the composition of the ferromagnetic alloy layer. The structural and compositional data are consistent with the shift of the $0-\pi$ transition toward the behaviour of heterostructures with different F layers. An important emerging indication to the physics of SFS is the weak relevance of the ideality of the interfaces: even in aged samples, with less--than--ideal interfaces, the temperature--induced $0-\pi$ transition is still detectable albeit at a different critical F thickness.
\end{abstract}
\maketitle

\section{Introduction}
\label{intro}
Ferromagnetism and singlet superconductivity are two antagonistic long--range orderings. It was demostrated that, however, they can coexist in Superconductor/Ferromagnet (SF) heterostructures. When and how this coexistence takes place, it has been extensively theoretically described in Ref.s \onlinecite{buzdinreview,golubovRMP}. In SF heterostructures a superconductor is in contact to a ferromagnet, and Cooper pairs can penetrate into the F layer, inducing an exponentially decreasing superconducting order parameter in the F region. The length scale is dictated by the ferromagnetic coeherence length $\xi_F$, due to the finite lifetime  of the Cooper pairs in the exchange field of the ferromagnet. In addition, the diffusion of the ferromagnetic ordering into the S layer(s) results in a decrease of the superconducting transition temperature $T_c$ (inverse proximity effect) of the whole structure. Moreover, the interaction between the exchange field and the Cooper pairs produces an oscillatory behaviour of the superconducting order parameter in the F region with a spatial periodicity, again on the scale of $\xi_F$. The damped oscillatory behaviour of the superconducting order parameter in the ferromagnetic region is responsible for a nonmonotonic dependence of $T_c$ on the thickness of the ferromagnetic layer.

In the specific, and much studied\cite{ryazanovPRL2001,kontosPRL2002,sellierPRB2003,obozonovPRL2006,pompeoPRB14} case of SFS trilayers, the competition between S and F orderings gives rise to two possible superconducting ground states, so--called ``$0$'' and ``$\pi$'' states. In the $0$ state the superconducting order parameter is depressed in the F layer, but no phase difference exists between the S layers. In the $\pi$ state the superconducting wavefunction shows a $\pi$ phase difference between the S layers, so that the order parameter crosses zero in the F layer. Whether the ground state in a SFS structure is the $0$ or $\pi$, it depends on $\xi_F$ (i.e., on the F material) and on the ferromagnetic layer thickness $d_F$.\cite{buzdinreview} For the very nature of the transition, the main experimental evidence of the $0-\pi$ transition has been gained through Josephson--like experiments.\cite{ryazanovPRL2001,sellierPRB2003,obozonovPRL2006} It is now widely accepted that, by varying the F thickness or the F material, one can observe the $0-\pi$ transition. 

A second, specific aspect of the $0-\pi$ transition concerns its true thermodynamic nature. Since the signature of the transition is observed or predicted mostly in phase--interference effects, one might raise some questions about the true nature of the transition. Recently, it has been experimentally observed\cite{pompeoPRB14} that the transition has a true thermodynamic nature: surface impedance measurements yielded the temperature dependence of the superfluid density in SFS samples with different $d_F$. In the sample with critical $d_F$ thickness, a reentrant jump of the superfluid density was observed, in agreement with theoretical calculations.%
%
{The experimental behavior was in agreement with theoretical calculations,\cite{pompeoPRB14,samokhvalovPRB2015} and the extended theoretical treatment\cite{samokhvalovPRB2015} confirmed the existence of a jump in superfluid density at the $0-\pi$, temperature--induced transition.%
}%
However, while all qualitative features of the theoretical calculations were observed in the measurements of the superfluid density, a few minor quantitative discrepancies existed. As an example, the temperature--induced transition occurred at a ratio $d_F/\xi_F$ different (by a factor $\sim$2) than predicted. At a previous stage, it was not possible to ascertain which aspects of the theory (such as, e.g., the perfect transparency of S--F barriers) were essential to the phenomenon. Thus, continuing the study of the SFS heterostructures is relevant to foster the understanding of the fundamental physics that makes the $0-\pi$ transition observable.

This paper presents a study of the electrodynamics, of the local structure, and of the compositional profile, of SFS, Nb/PdNi/Nb, heterostructures, successive to a 2 years ageing, in order to give information on the $0-\pi$ transition after possible deterioration of the samples.%
{
The ageing of the samples helps to stress the effect of structural properties, such as the sharpness of the interfaces ordefects in the  composition, which may be crucial to quantitatively match the theory and the experimental evidence.%
}%
The paper is organized as follows: in Sec.\ref{samples} we summarize the preparation of the SFS heterostructures; in Sec.\ref{mw} we present microwave electrodynamics in aged samples, as compared to fresh samples; in Sec.\ref{XAS} we present the results of extended X--ray absorption fine structure (XAFS) measurements carried out at the Nb, Pd and Ni $K$--edges, yielding detailed information on the local structure around each species in our samples; in Sec.\ref{TOF} we explore the compositional depth profile of our samples, with 1 nm resolution, obtaining clear information on the interfaces and on possible interdiffusions; finally, in Sec \ref{disc} we discuss the results, and give short conclusions.

\section{Samples and initial characterization}
\label{samples}

Nb/Pd$_{84}$Ni$_{16}$/Nb SFS trilayers have been grown at room temperature on Al$_2$O$_3$ (1120) substrates by ultra-high-vacuum dc diode magnetron sputtering. The base pressure in the chamber was in low 10${-8}$ mbar, while the Argon pressure during the deposition was $P_{Ar}$=3 $\times$10${-3}$ mbar. The SFS trilayers consisted of two Nb layers of nominal thickness $d_S=15$ nm, sputtered at a rate of  $r_S=0.28$ nm/s, and an intermediate Pd$_{84}$Ni$_{16}$ one of different thickness $d_F=$ 2--10 nm, deposited at a rate of $r_S=0.40$ nm/s. The deposition rates were calibrated by independent X-Ray Reflectivity measurements on deliberately deposited single layers of each material sputtered for a given deposition time. Samples are labeled by the letter $d$ followed by a number indicating the thickness of the Pd$_{84}$Ni$_{16}$ in nm (e.g., ``sample d8'' is the trilayer with $d_F=$ 8 nm). A pure Nb film 30 nm thick was also grown for comparison. More details concerning the sample preparation are described in Ref. \onlinecite{SUST24}. Pd$_{1-x}$Ni$_{x}$ is a weak ferromagnetic alloy widely employed in the framework of the research of the coexistence of S and F orderings, due to the tunability of its exchange energy $E_{ex}$ by a small amount of the Ni content. In particular, the magnetic and electrical properties of Pd$_{84}$Ni$_{16}$ (hereafter PdNi) were widely investigated in Ref.\onlinecite{SUST24}, resulting in an exchange energy $E_{ex}\simeq 14$ meV, a ferromagnetic coherence length $\xi_F\simeq 3$ nm and a Curie temperature $T_{c}=200$ K. Moreover, it is also worth noticing that a number of works present in the literature suggest that Nb/PdNi-based systems show relatively high transparency of the S/F interface.\cite{Cirillo2005}

\section{Superfluid density measurements}
\label{mw}
\begin{figure}[ht]
\centering
\includegraphics  [width=.8\columnwidth]{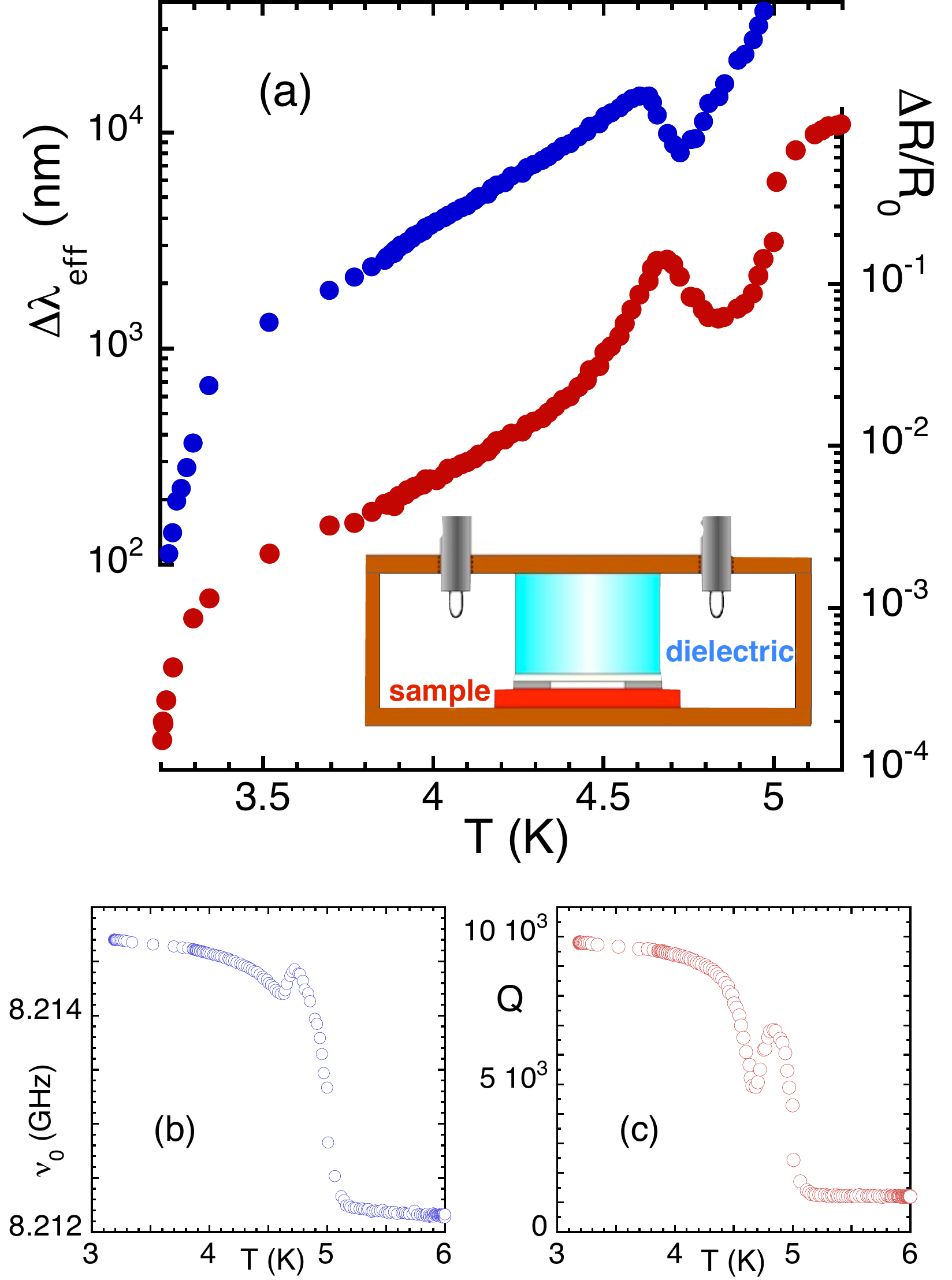}
\caption{(a) measurements on sample d2 showing the reentrant behaviour, signature of the $0-\pi$ transition:\cite{pompeoPRB14} $\Delta\lambda_{eff}(T)$ (blue full dots), and $\Delta R_s(T)/R_0$ (red full dots). Inset: sketch of the dielectric resonator and of the film position. The reentrant behaviour is visible already in the raw data for $\nu_0(T)$ (b) and $Q(T)$ (c). Reference temperature is $T_{ref}=$3.19 K.}
\label{figexp} 
\end{figure}

Microwave measurements are a useful tool to investigate the fundamental properties of superconducting materials, since they give information on both quasiparticles and superfluid density. We measured the microwave response by means of a cylindrical dielectric resonator, operating at $\sim$8 GHz. A sketch of the experimental setup is reported in the inset of Fig.\ref{figexp}a. The details of the setup and of the procedure have been given elsewhere.\cite{pompeoMSR14,Torokhtii-PhC12}
\begin{figure}[ht]
\centering
\includegraphics  [width=.8\columnwidth] {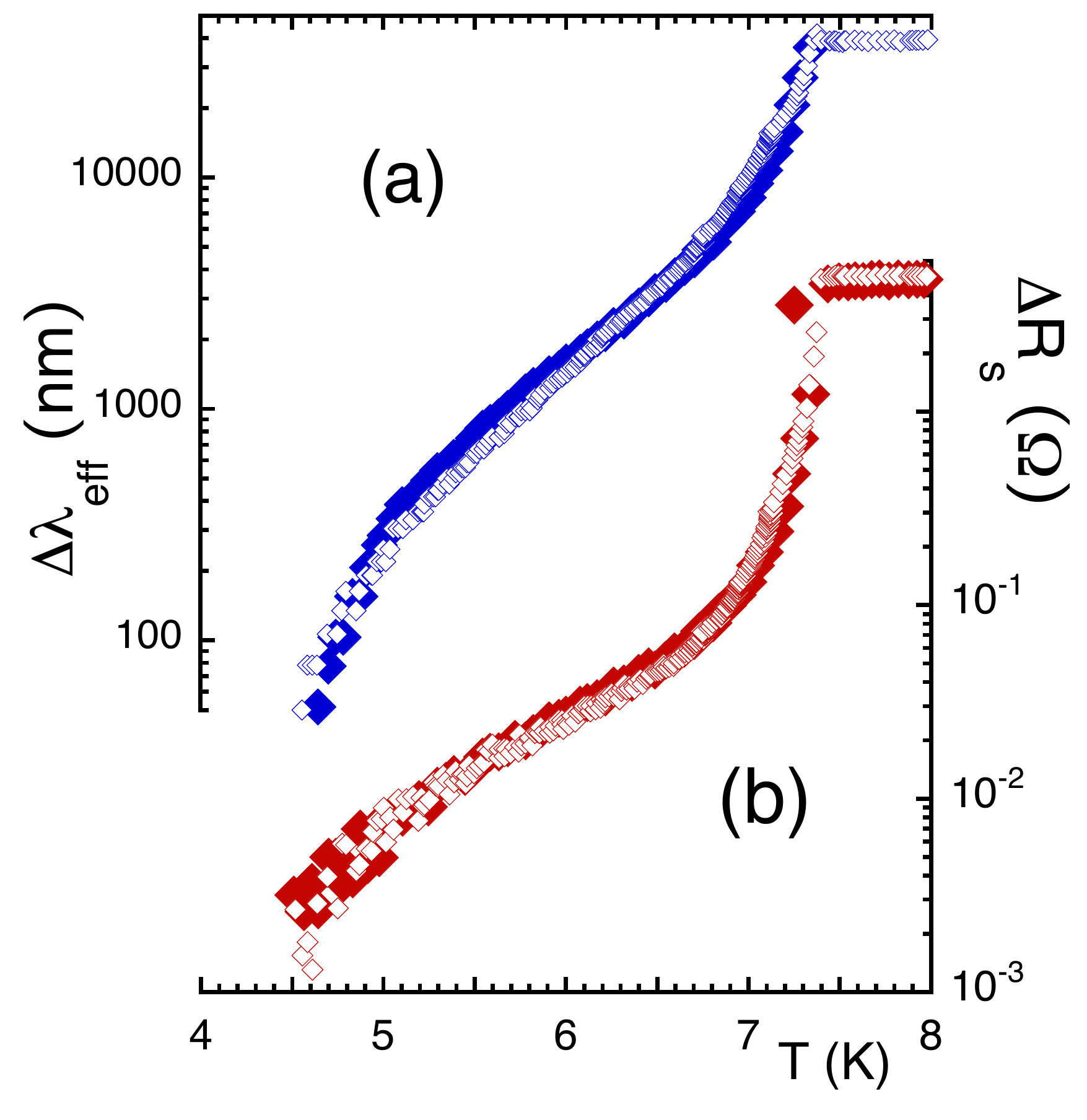}
\caption{Microwave response of the fresh and aged pure Nb sample. Open symbols: fresh sample. Full symbols: aged sample. (a) $\Delta\lambda_{eff}$; (b) $\Delta R_s$. The reference temperature is $T_{ref}=$~4.50~K. The flattening of the data indicates that the resonator reached the sensitivity limit, and it is not a direct measure of the normal state.}
\label{figNb} 
\end{figure}
In short, the sample was placed at the bottom of the resonator. Measurements of the quality factor $Q(T)$ and of the resonant frequency $\nu_0(T)$ yielded the variation of the effective surface resistance $\Delta R_s$ and of the effective penetration depth $\Delta\lambda_{eff}$ with $T$, according to:
\begin{eqnarray}
\hspace{1cm}
\Delta\lambda_{eff}(T)= \lambda_{eff}(T)-\lambda_{eff}(T_{ref})=\\
\nonumber
= - \frac{G_s}{\pi\mu_0}\frac{\nu_0(T)-\nu_0(T_{ref})}{\nu_0^2(T_{ref})},
\label{eq:deltalam}
\end{eqnarray}
and
\begin{eqnarray}
\hspace{1cm}
\Delta R_{s}(T)=R_s(T)-R_s(T_{ref})=\\
\nonumber
=G_s\left(\frac{1}{Q(T)}-\frac{1}{Q(T_{ref})}\right).
\label{deltaRs}
\end{eqnarray}
where $T_{ref}$ is a reference temperature and $G_s$ is a geometrical factor that can be computed by standard simulation software. The most direct information is given by Eq.\eqref{eq:deltalam}: in fact, in thin structure of overall thickness $d$ as ours, and not too close to $T_c$, one has:\cite{silvaSUST96,pompeoSUST05}
\begin{equation}
\hspace{1cm}
\lambda_{eff}(T)= \lambda(T)\coth\frac{d}{\lambda(T)}\approx\frac{\lambda^2(T)}{d}
\label{eq:lameff}
\end{equation}
so that\cite{tinkham} $\lambda_{eff}(T)\propto 1/n_s(T)$, with $n_s(T)$ the superfluid density. Thus, Eq.\eqref{eq:deltalam} directly links the measured parameter $\nu_0(T)$ to the thermodynamic property $n_s(T)$.
\begin{figure}[ht]
\centering
\includegraphics  [width=.75\columnwidth] {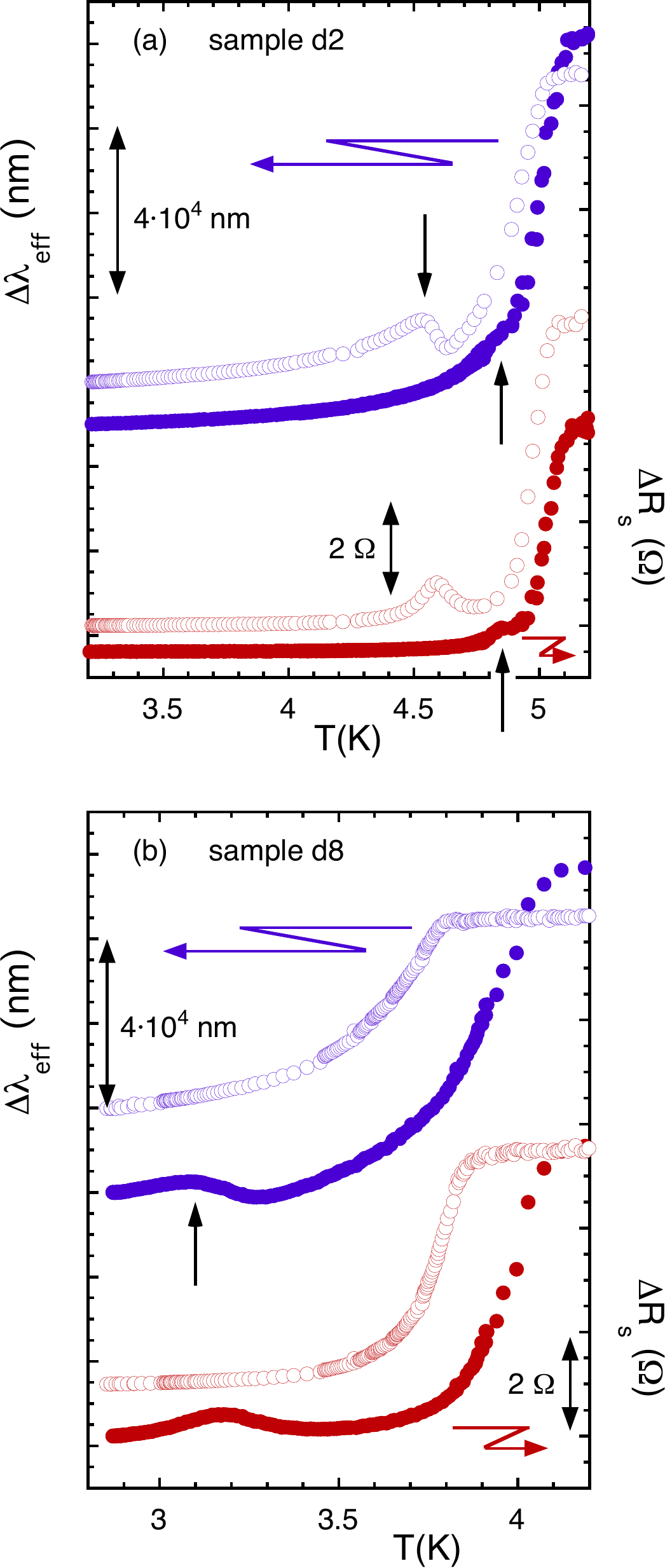}
\caption{Comparison of the electrodynamics response of the fresh (open symbols) and aged (full symbols) SFS trilayers. Blue (red) symbols stand for $\Delta\lambda_{eff}$ ($\Delta R_s$). Data sets have been offset for clarity. (a) sample d2; $T_{ref}=$~3.19~K. The arrows mark the peak of $\Delta\lambda$, corresponding to the reentrant jump of the superfluid density at the $0-\pi$ transition, before (downward arrows) and after (upward arrows) ageing. $\Delta R_s$ behaves consistently. (b) sample d8, same symbols as in (a); the arrow marks the appearance of he signature of the $0-\pi$ transition after ageing; note also the upward shift of $T_c$ after ageing.}
\label{figmwaged} 
\end{figure}

In a previous work\cite{pompeoPRB14} we have measured $\Delta\lambda_{eff}(T)$ in samples d2, d8, d9, and in the Nb sample as a reference. The presence of a sharp {\em increase} of $\Delta\lambda_{eff}$ (that is, a {\em decrease} in $n_s$) with decreasing temperature, at $T_{0\pi}<T_c$, was the evidence for the temperature--induced $0-\pi$ transition, from the high--$T$, $0$ state, to the low--$T$, $\pi$ state. Notably, it was possible to observe the $T$--induced transition only in the sample with the specific F thickness $d_F=$ 2 nm. Interestingly, above this thickness a departure from conventional behaviour was observed in the mixed--state properties, such as the flux--flow resistivity,\cite{Silva-SUST11,TorokhtiiJS13} the vortex pinning constant,\cite{torokhtiiMSMW,pompeoJS15} and the flux--creep factor.\cite{Pompeo-JOSC13} Some of these observations were related to a change in the superconducting state. Thus, there is evidence that the $0-\pi$ transition marks a boundary between different superconducting states, with the $\pi$ state firmly established in samples with large enough $d_F$ (that is, with the superconducting transition directly into the $\pi$ state), and a double transition (in the superconducting $0$ state at $T_c$, and then in the $\pi$ state at $T_{0\pi}<T_c$) in the sample with critical F thickness.

In order to show the experimental signature of the $T$--induced $0-\pi$ transition, data for $\Delta\lambda_{eff}(T)$ and for $\Delta R_s(T)$, as well as the corresponding raw data for $\nu_0(T)$ and $Q(T)$ are reported in Fig.s \ref{figexp}b and \ref{figexp}c. The data refer to ``fresh'' samples, measured within a few weeks after preparation. We emphasize that the flattening of the data above $T_c$ is due to the sensitivity limit of the resonator: the flat value should not be taken as the ``normal state'' value.%
{
Since we have identified the nonmonotonous behavior of $\Delta\lambda_{eff}(T)$ as the signature of the $0-\pi$ transition, it is important to rule out possible spurious effects that can mimic this behavior. An extensive discussion has been reported previously,\cite{pompeoPRB14} we shortly recall it here. First, several experimental runs were performed, including runs after entirely disassembling and reassembling the microwave resonator: the data completely overlapped. Second, the same jump in $\Delta\lambda_{eff}(T)$ obtained by direct measurements in zero magnetic field  (as those here reported), was obtained by estimating the effective penetration depth from the condensation energy derived from magnetic--field--dependent microwave measurements (a completely different experiment): the results from the different experiments were in excellent agreement. Moreover, multiple transitions are ruled out by the fact that they would lead to a stepwise temperature dependence of $\Delta\lambda_{eff}(T)$, and not to a reentrance (note that the data for  $\Delta R_s(T)$ point to the same conclusion). A more subtle spurious effect may reside in the stray field of the magnetic layer itself and/or of magnetic small particles embedded in the superconducting matrix. Such phenomena have been widely studied,\cite{stamopoulosPRB2005,stamopoulosPRB2007} with the largest influence on the magnetoresistance and the magnetization in external magnetic fields. We focus here on the few effects that can be observable in our zero--field measurements. In superconducting films with different local concentration of ferromagnetic nanoparticles, a multi--state resistance value could be observed.\cite{stamopoulosPRB2005} The effect was enhanced by an external magnetic field (which is not our case), and most importantly it would similarly give, in our sample--averaged measurements, a multi--step transition both in $\Delta R_s(T)$ as well as in $\Delta\lambda_{eff}(T)$. The effect would be similar to the transition of a multiphase superconductor, while the observed behavior is a reentrance of the superconducting properties. A second possible effect comes from the stray field due to the coupling of the external ferromagnetic layers in FSF heterostructures:\cite{stamopoulosPRB2007} the stray field could locally exceed the critical field. In our SFS heterostructures, that contain a single F layer, this phenomenon is likely to be much weaker. Moreover, the effect on our sample--averaged measurements would again be similar to a multiphase superconductor, thus giving rise to a multi--step or a broadened transition. Thus, we believe that spurious effects are unlikely to produce a reentrance in the superconducting propeorties, in our experimental configuration and with the check we have made.%
}%

As it will be largely discussed in the following, we anticipate that repetition of the measurements on the same samples after 24 months yielded strongly different results (while short--term repetition of the measurements gave identical results, as discussed in Ref.\onlinecite{pompeoPRB14}). Since aim of this paper is to understand this ageing effect and gather informations from joint electrodynamics, structural and chemical investigation, we performed a preliminary repetition of the measurements on the reference Nb sample, as--grown and after ageing. The results are presented in Fig. \ref{figNb}, in terms of $\Delta\lambda_{eff}(T)$, Fig.\ref{figNb}a, and  $\Delta R_s(T)$, Fig.\ref{figNb}b. As it can be seen, after a rather long time of 24 months the measurements are practically identical. Thus, we conclude that any ageing effect that should be observed must not be ascribed to any deterioration of the Nb itself, but only to alterations of the PdNi layer or of the F/S interfaces

Fig. \ref{figmwaged} reports the main experimental findings of this Section. There, we report and compare $\Delta\lambda_{eff}(T)$ and $\Delta R_{s}(T)$ obtained for the fresh and aged trilayers d2 and d8. Several effects emerge. First, $T_c$ of the sample d2 is little changed (a small shift upward, $\Delta T_c \lesssim $~50~mK,  might be inferred), while in sample d8 $T_c$ clearly shifts {\em upward} with ageing: this result is in clear contradiction with any na\"ive interpretation of the ageing effects in terms of just a deterioration of the samples. The involved phenomena are clearly more complex.

The second important effect is the shift at higher $T$ of the $0-\pi$ transition in sample d2, where it is accompanied by a broadening and weakening, as indicated by the very faint remaining features, and the emergence of the fingerprint of the transition in sample d8, although in a broader fashion than in the fresh sample d2.

It is then clear that the ageing produce changes in the PdNi layer, that affect the magnetic properties and then the overall superconducting order parameter, in a nontrivial fashion. Thus, before any discussion of the electrodynamics data a structural and compositional analysis is in order. The following Sections are devoted to this task.

\section{X-ray absorption spectroscopy}
\label{XAS}

X-ray absorption spectroscopy (XAS) is a chemical selective probe\cite{XAS} that we used to investigate the local structure around Nb, Ni and Pd atoms in the aged samples and to individuate the modifications of the local structure and of the composition with respect to the nominal parameters. To this aim we have collected Nb $K$--edge XAFS spectra at the ESRF--BM08 beamline\cite{GILDA} (Grenoble, France), and Pd and Ni $K$--edge XAFS spectra at the ELETTRA XAFS\cite{ELETTRA} beamline (Trieste, Italy). 
XAS spectra were measured in fluorescence geometry, keeping the samples at the liquid nitrogen temperature in order to reduce the thermal contribution to structural disorder. Nb $K$--edge XAS spectra were collected using an ultrapure Ge multidetector (13 elements), Pd and Ni $K$--edge XAS spectra were measured using a KETEK silicon drift detector.
\begin{figure*}[ht]
	\begin{minipage}[t] {0.35\columnwidth}
	\raggedright
 	\par\vspace{0pt}
	\includegraphics[angle=0,height=3cm]{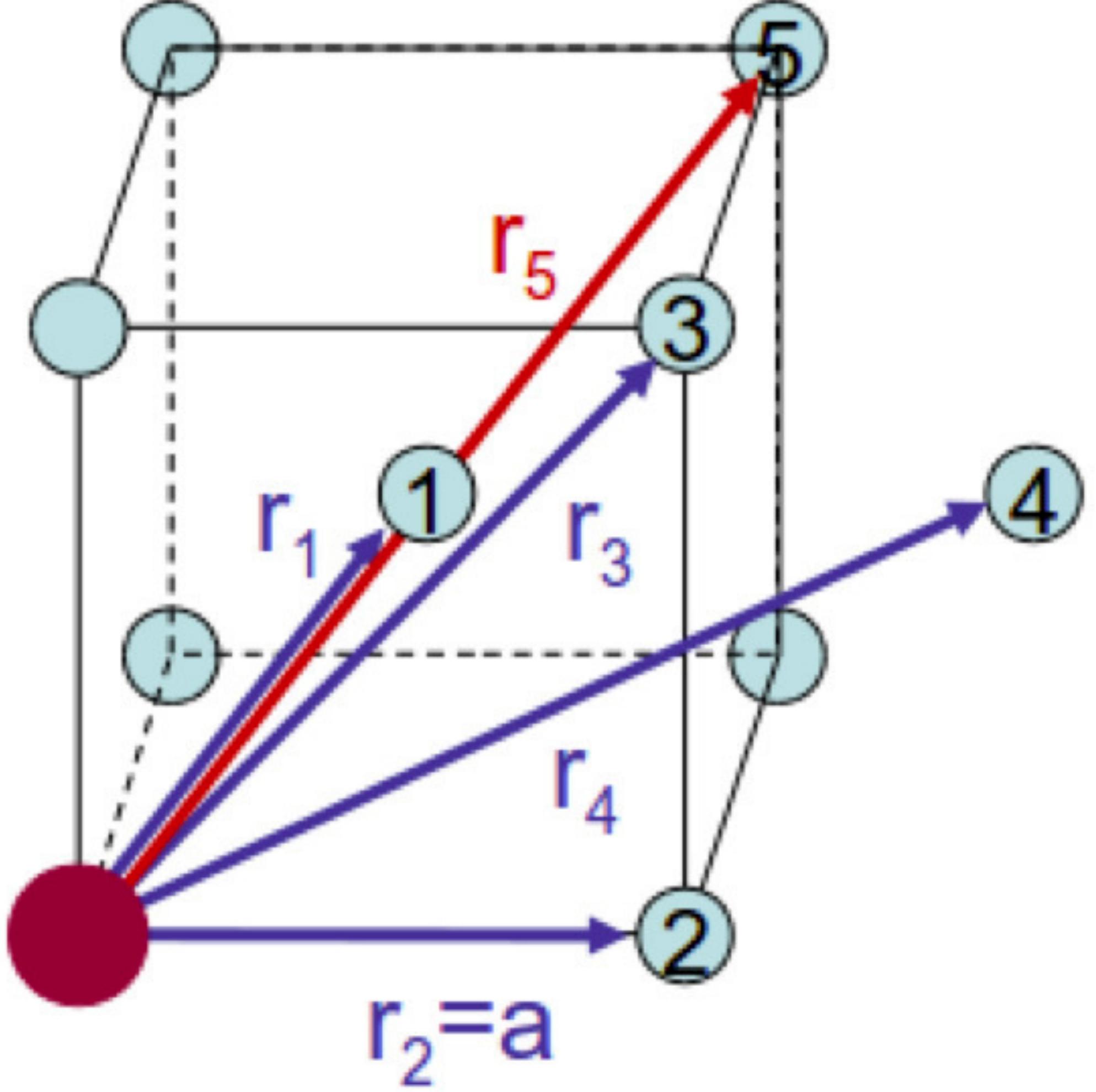}\\
	(a)
	\end{minipage}
	\begin{minipage}[t] {0.5\columnwidth}
	\raggedright
 	\par\vspace{0pt}
 	\small
	\begin{tabular}[c]{|c|c|c|}	
		\hline
		Path & Distance & Coordination \\     
		 &  &  number \\     
		\hline
		I      & $r_1=a\sqrt{3}/2$  &  8\\
		\hline
		II      & $r_2=a$  &  6\\
		\hline
		III     & $r_3=a \sqrt{2}$  &  12\\
		\hline
		IV    & $r_1=a \sqrt{11} / 2$  &  24\\
		\hline
		V      & $r_1=a \sqrt{3}$  &  8\\
		\hline
	\end{tabular}
	\end{minipage}
	\hspace{1cm}
	\begin{minipage}[t] {0.35\columnwidth}
	\raggedright
 	\par\vspace{0pt}
	\includegraphics[angle=0,height=3cm]{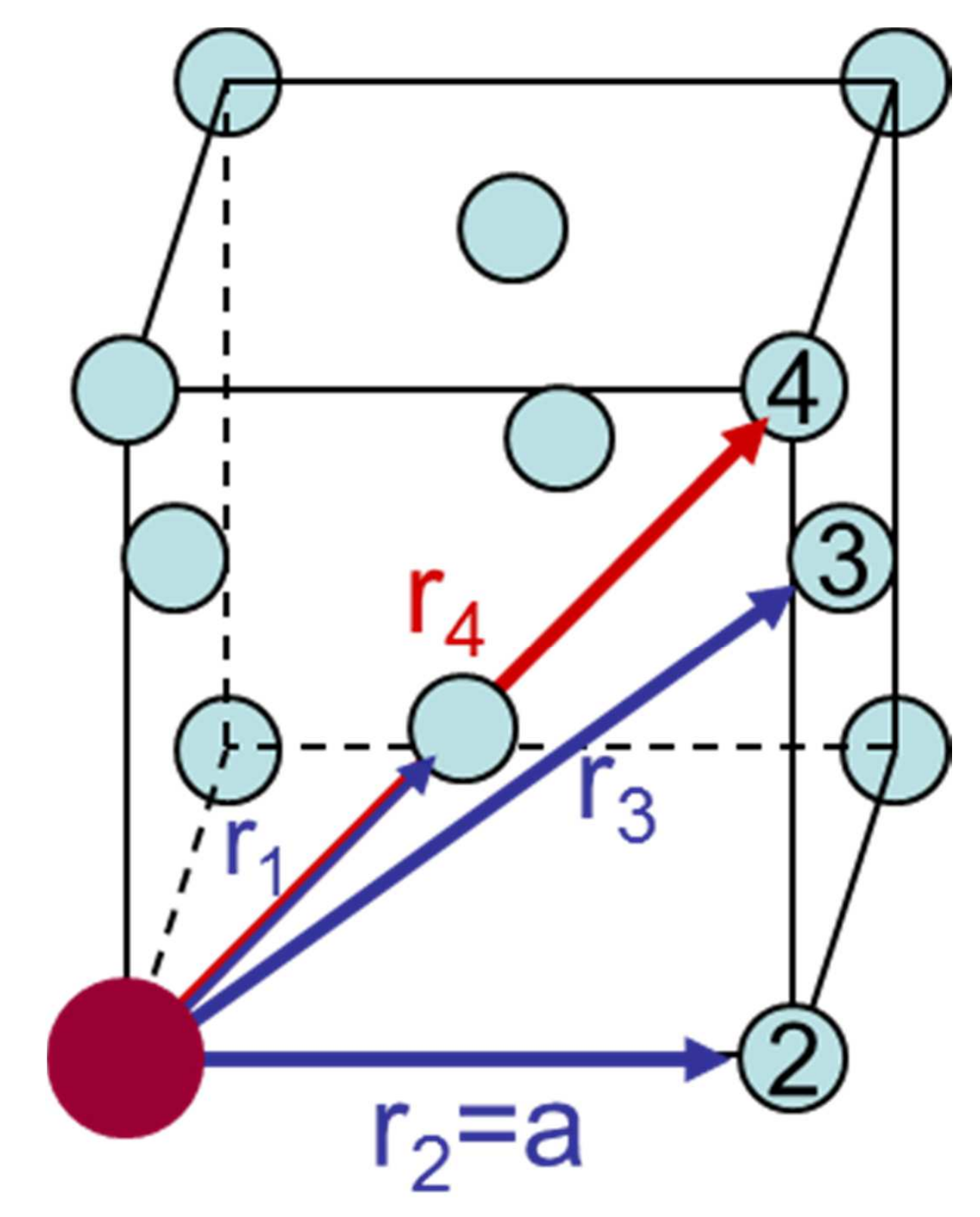}\\
 	(b)
	\end{minipage}
	\begin{minipage}[t] {0.5\columnwidth}
	\raggedright
 	\par\vspace{0pt}
	\small
	\begin{tabular}[c]{|c|c|c|}
		\hline
		Path & Distance & Coordination \\     
		 &  &  number \\   
		\hline
		I      & $r_1=a\sqrt{2}/2$  &  12\\
		\hline
		II      & $r_2=a$  &  6\\
		\hline
		III     & $r_3=a \sqrt{6}/2$  &  24\\
		\hline
		IV    & $r_1=a \sqrt{2}$  &  12\\
		\hline
	\end{tabular}
	\end{minipage}
\caption{(a) Scattering paths for the bcc crystal structure of Nb and (table) distances and coordination numbers of the first five shells; (b) Scattering paths for the fcc crystal structure and (table)  distances and coordination numbers of the first four coordination shells. The XAFS data analysis only considered the first two coordination shells.}
\label{fig:strutt}
\end{figure*}
%
In order to achieve suitable data statistics and remove the Bragg peaks from the substrate, several spectra were measured for each sample (three for Nb $K$--edge and at least five for Pd and Ni $K$--edge) with a slight rotation of the samples at each scan (approximately $\pm$ 1 deg.) to allow displacing the Bragg peak positions on the spectra.\cite{V2O3} This allowed to recognize and remove these peaks. The cleaned absorption spectra were then averaged up for the analysis.

\begin{figure}[h]
	\begin{minipage} {1\columnwidth}
	\centering
	\includegraphics  [width=0.7 \columnwidth] {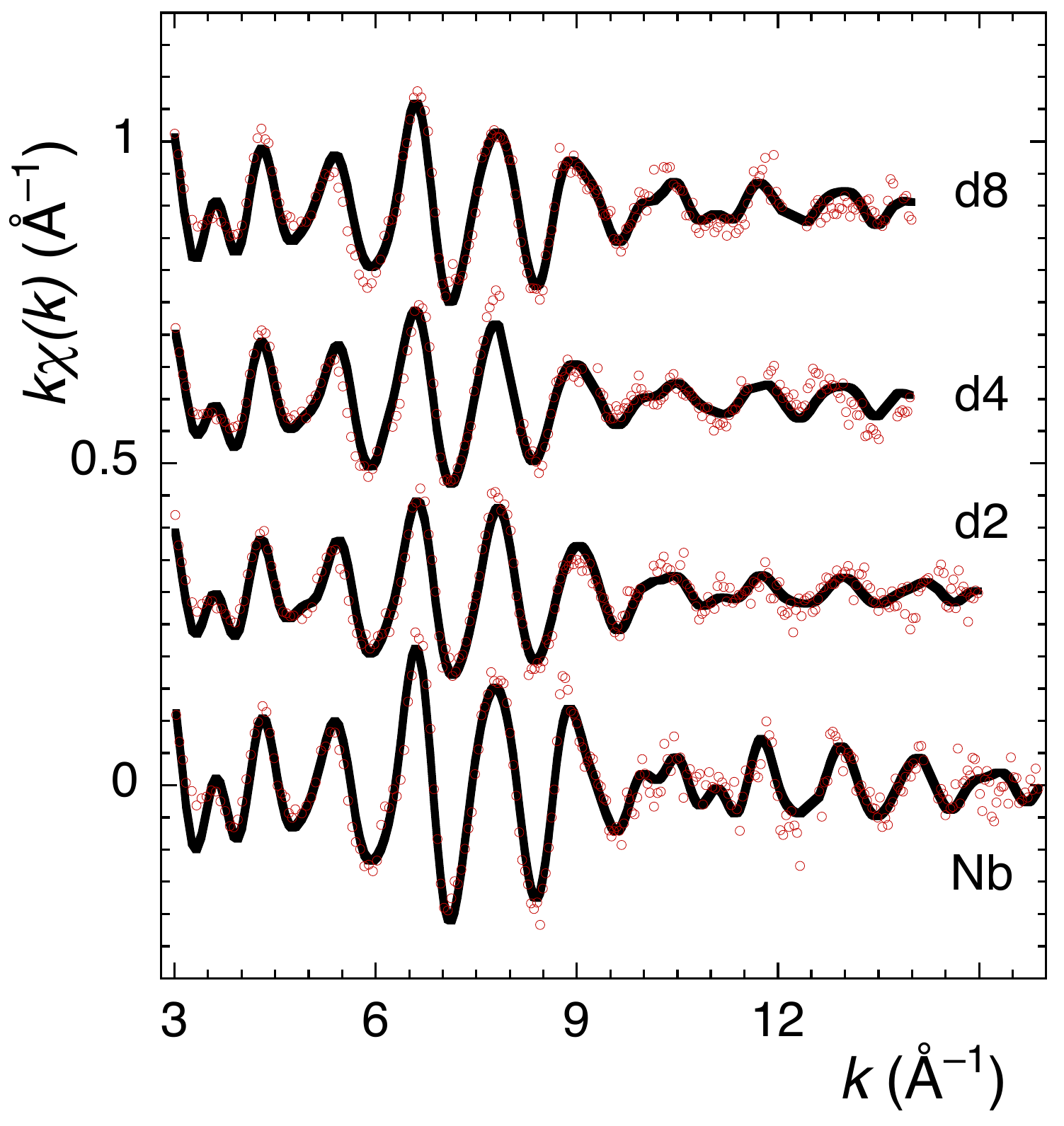}
%
%
\small
\begin{center}
\begin{tabular}{|ccccc|}
\hline
 & \multicolumn{2}{c} {Shell I (N=8)}& \multicolumn{2}{c} {Shell II (N=6)}\vline\\[-1ex]
\raisebox{1.5ex}{Sample} & $R_I$ (\AA) & $\sigma^2 \times10^3 ($\AA$^2$) & $R_{II}$ (\AA)& $\sigma^2 \times10^3$(\AA$^2$) \\[0.5ex] 
\hline
Nb      & 2.86  &  5.4(4)  & 3.30(1) & 4.0(4)\\
\hline
d2 & 2.85  &  7.7(6)  & 3.29(1) & 8.7(7)\\
\hline
d4 & 2.87  &  7.3(6)  & 3.31(1) & 6.8(7)\\
\hline
d8 & 2.85  &  6.6(6)  & 3.29(1) & 6.8(7)\\
\hline
\end{tabular}
\end{center}
\end{minipage}
\caption{$k$--weighted EXAFS data (red points) and best fit (full black lines) for the Nb $K$--edge data. Table: fit parameters for the interatomic distances and the Debye-Waller factor for the first two coordination shells.%
{
Note that $R_{II}=a_{Nb}$.%
}
The statistical errors on the refined parameters are shown in parenthesis,  the values for $R_I$ are constrained to be $a_{Nb}\sqrt{3}/2$.}
\label{figNbspectra} 
\end{figure}

\begin{figure}
	\centering
	\includegraphics[width=0.7\columnwidth]{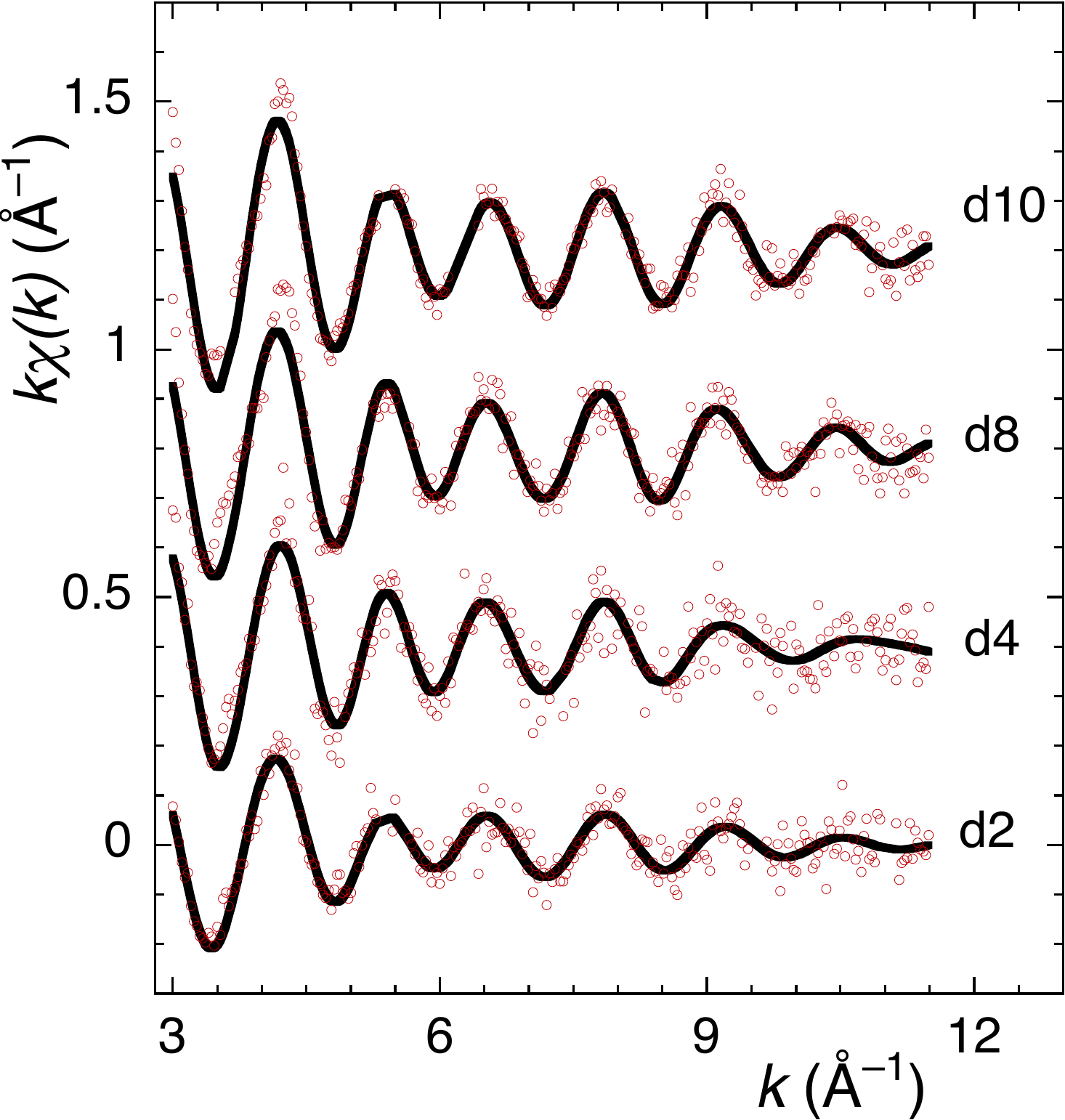}\\
	\vspace{0.5cm}
	\centering
	\small
		\begin{tabular}[c]{lllll}
                     \multicolumn{5}{c} {Ni $K$--edge}\\
		\toprule
		 & \multicolumn{3}{c} {PdNi}\\
		\raisebox{1.5ex}{Sample} & $a_{PdNi}$ &$\sigma^2_I\times10^3$ & $\sigma^2_{II}\times10^3$ &   $x$\\[0.5ex]
                                 &      (\AA{})     &  (\AA$^2$)&  (\AA$^2$) &  (\% ) \\
		\midrule
		d2      &       3.80*     & 8(3)          & 19(3)      & 65(5)    \\
		d4      &       3.84*     & 11(3)        & 14(3)      & 80(6)    \\
		d8      &       3.83*     & 6.2(3)       & 12(3)      & 76(4)    \\
		d10    &       3.82*     & 5.3(3)       & 16(3)      & 72(4)    \\
		\bottomrule
		\toprule
		 & \multicolumn{3}{c} {Ni}\\[.5ex]
		\raisebox{1.5ex}{} & $a_{Ni}$  &      $\sigma^2_I\times10^3$ & -  &  $Y$     \\[0.5ex]
                                &      (\AA{})     &  (\AA$^2$)&   &  (\% ) \\
		\midrule
		d2      &    3.50(2) &     2.0**           &    &19(3)  \\
		d4      &    3.51(2) &     2.0**           &    & 8(2)   \\
		d8      &    3.50(2) &     2.0**           &    & 8(2)   \\
		d10    &    3.50(2) &     2.0**           &    & 9(2)   \\
		\bottomrule
		\end{tabular}
\caption{$k$--weighted XAFS data at the Ni edge (coloured dots) and best fits (full black lines). Parameters are reported in the table. It is found that Ni does not bond to Pd only, but it is also found close to Nb, in the NiNb$_5$ alloy.\\
** In PdNi, $a_{PdNi}$ is not a free parameter (see text);\\
*** In the Ni--rich phase, $\sigma^2_I$ is kept constant (see text).}
\label{figNiEXAFS}
\end{figure}

\begin{figure}
	\centering
	\includegraphics[width=0.7\columnwidth]{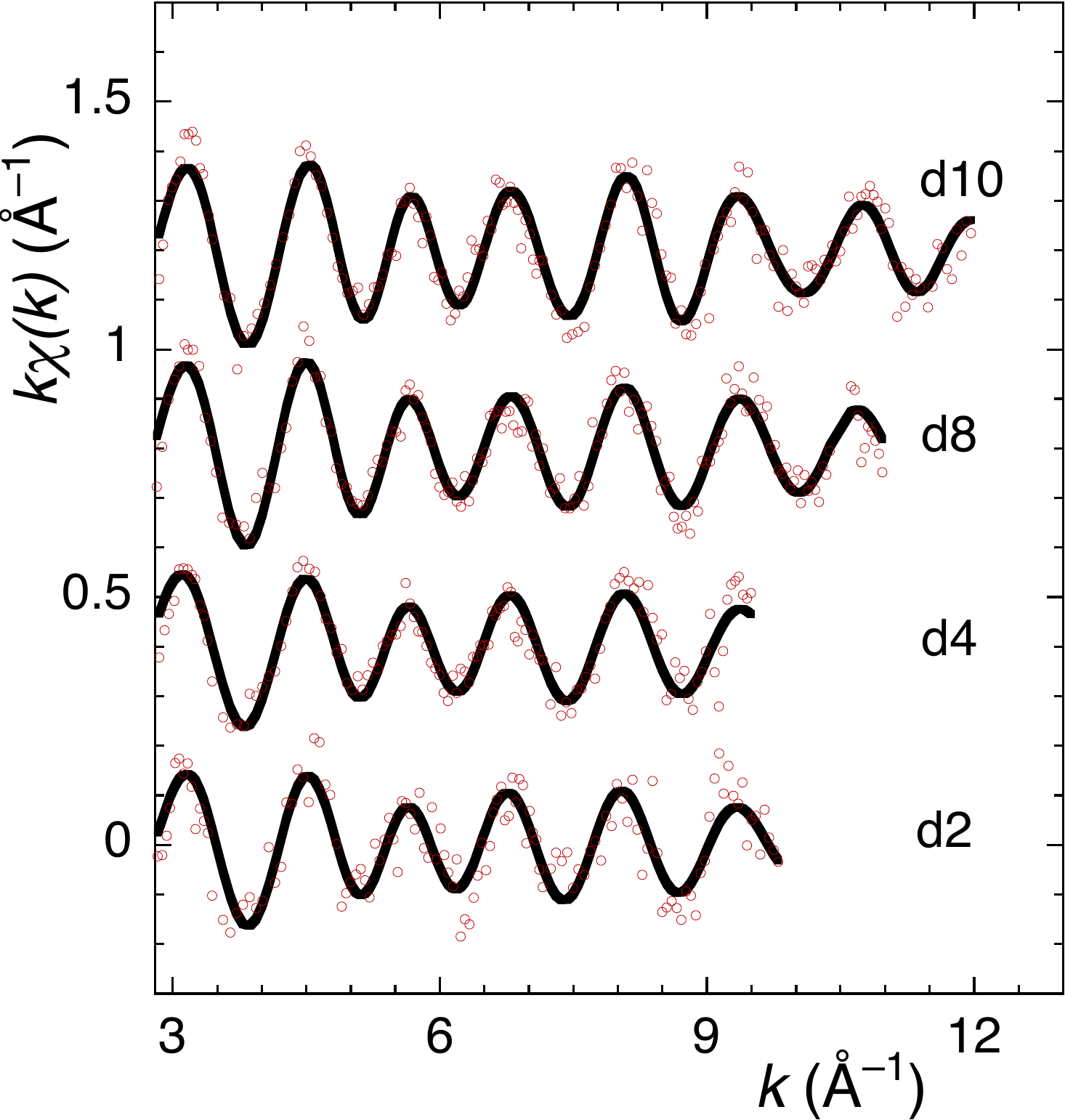}\\
	\vspace{0.5cm}
	\centering
	\small
		\begin{tabular}[c]{lllll}
                     \multicolumn{5}{c} {Pd $K$--edge}\\
		\toprule
		 & \multicolumn{3}{c} {PdNi}\\
		\raisebox{1.5ex}{Sample} & $a_{PdNi}$ &$\sigma^2_I\times10^3$ & $\sigma^2_{II}\times10^3$ &   $x$\\[0.5ex]
                                 &      (\AA{})     &  (\AA$^2$)& (\AA$^2$) &  (\% ) \\
		\midrule
		d2      &       3.83*       & 6.4(5)         &15(6)        & 76(4)     \\
		d4      &      3.82*        &  5.2(4)        &12(6)        & 72(5)    \\
		d8      &      3.84*        &  4.4(4)        & 12(5)       & 80(3)    \\
		d10    &       3.83*      & 4.2(3)        & 4.1(3)       & 76(3)    \\
		\bottomrule
		\toprule
		 & \multicolumn{3}{c} {Nb$_3$Pd}\\[.5ex]
		\raisebox{1.5ex}{} & $a_{NbPd}$ (\AA{})  &      $\sigma^2_I\times10^3$ &  $\sigma^2_{II}\times10^3$ &  $Y$     \\[.5ex]
                                &      (\AA{})     &  (\AA$^2$) & (\AA$^2$) &  (\% ) \\
		\midrule
		d2      &    3.19(2)  &  11**            &  16**     & 65(5)  \\
		d4      &    3.17(1)  &  11(4)        &   16(5)   & 60(5)  \\
		d8      &    3.16(1)  &  12(4)        &   15(4)   & 50(3)  \\
		d10    &    3.16(1) &  9(3)           &   13(4)   & 51(3)   \\
		\bottomrule
		\end{tabular}
\caption{$k$--weighted XAFS data at the Pd edge (coloured dots) and best fits (full black lines). Parameters are reported in the table. It is found that Pd does not bond to Ni only, but it is also found close to Nb, in the Nb$_3$Pd alloy.\\
** In PdNi, $a_{PdNi}$ is not a free parameter (see text).
}
\label{figPdEXAFS}
\end{figure}

The structural EXAFS signal has been extracted following the standard procedures for pre-edge background subtraction, post edge ($\mu_0$) spline modelling and normalization.\cite{EstraFitexa} The EXAFS data were analysed by applying a multi-shell model taking into account for next neighbour distribution, including the relevant multiple scattering contributions. The inclusion of the next neighbour shells allows to reliably recognize the crystallographic phases hosting the absorber, which is a key information in this work. Constraints among the refined parameters were applied, based on crystallographic models.\cite{Battocchio} The experimental $k$-weighted $k\chi(k)$ EXAFS spectra were fitted to the  theoretical curves in the reciprocal space ($k$--space) (Fig.s \ref{figNbspectra},\ref{figNiEXAFS},\ref{figPdEXAFS}). Theoretical curves are calculated using the standard EXAFS formula in the Gaussian approximation.\cite{XAS}
The photoelectron scattering amplitude and the phase shift functions and the photoelectron mean free path were calculated ab-initio using FEFF code\cite{FEFF} for atomic clusters based on bulk structure.

The trilayer structure is  nominally  made of a Pd$_{84}$Ni$_{16}$ layer of variable thickness $d_F = $~2$\div$10 nm, sandwiched between two Nb layers of thickness $d_S=$~15 nm. In the analysis we started using  the bulk structure of Nb and PdNi phases as models, other contributions were added considering the known crystallographic phases in order to take into account for interdiffusion effects (see below).  

We discuss first the Nb local structure. The Nb bulk structure is bcc (Fig.\ref{fig:strutt}a), with lattice paramenter $a_{Nb}= $~3.3 \AA, and multiplicity numbers and crystallographic distances (in terms of the crystallographic lattice parameter $a_{bcc}$) are summarized in the same Figure. 
In the present case, the neighbour distances and multiplicity numbers were constrained to the bcc structure. Then, only the lattice parameter, $a_{Nb}$, and disorder factors, $\sigma_j$, were refined. 
The data refinement proceeded progressively by adding additional coordination shells on the basis of their relative amplitude, and the improvement of the best fit was statistically tested.\cite{EstraFitexa} Satisfactory agreement was achieved using the first five coordination shells, allowing to reproduce the Nb local structure till about 5.7 \AA. Noticeably, the V coordination shell involves intense multiple scattering terms in the EXAFS signal due to the collinear atomic configuration along the cube diagonal\cite{XAS} (focusing effect). The statistical analysis shows that the addition of further contributions does not improve the best fit quality.

In Fig. \ref{figNbspectra} we report the experimental $k \chi(k)$ in all samples measured (red dots), together with the fits (continuous lines). The good fittings demonstrate that the Nb bcc bulk structure is a good model for the average structure around Nb in the S layers. 
The best fit parameters are in agreement with Nb bulk structure, in particular $a_{Nb}\simeq$ 3.3 \AA\ is unchanged within the uncertainty for all the investigated samples, demonstrating that the bcc bulk-like phase is preserved for the Nb layers in all the samples measured.


The EXAFS signal in multilayers appears attenuated with respect to pure Nb film (Fig.\ref{figNbspectra}). This finding must be attributed to a somewhat larger disorder of the Nb structure in the trilayers, suggesting some influence on the S layer by the F layers. The quantitative analysis (best fit parameters) shows that the Nb--Nb interatomic distances are unchanged in the trilayers. The structural disorder ($\sigma^2$ parameters) in the trilayers is higher than in pure Nb film, but it is similar in all aged trilayers (within the uncertainty), suggesting that the disorder induced in the Nb layers is weakly dependent on the PdNi film thickness.

Further understanding about the mechanisms giving rise to the structural disorder and to the modification of the superconducting response of the heterostructures can be obtained looking at the local atomic structure around Pd and Ni atoms.  

Ni and Pd $K$--edge XAS spectra were measured on 4 samples with increasing PdNi layer thickness $d_F$, from 2 to 10 nm. To gain more insight on the trend of the disorder around Pd and Ni, here we have extended the investigation to the sample with $d_F=$~10~nm.
Pd and Ni have the fcc structure (Fig.\ref{fig:strutt}b) with lattice parameters $a_{Pd}$=3.89\AA\ and  $a_{Ni}$=3.52\AA,\cite{Wyckoff} respectively.   Literature data reports that the Pd$_{x}$Ni$_{1-x}$ alloys have the fcc structure with  lattice parameters ranging from 
 $a_{x=0.5}$=3.76 \AA,\cite{Lihl} to
$a_{x=0.9}$=3.87 \AA.\cite{Flanagan}

The analysis of Pd and Ni $K$--edge spectra  is relatively complex and detailed with care in the appendix A, here we focus on the structural findings. 

The $k$--weighted Pd and Ni $K$--edge experimental EXAFS spectra $k\chi(k)$, along with the best fit curves, are shown in Fig.s \ref{figNiEXAFS} and \ref{figPdEXAFS} for all the four trilayers, and the resulting parameters for the best fits are reported in the tables accompanying Fig.s \ref{figNiEXAFS} and \ref{figPdEXAFS}. 
Pd XAFS spectra present higher statistical noise for thinner samples, in particular d2 and d4 samples had to be analysied in a restricted $k$--range. The amplitude of the EXAFS signal steadily increases with increasing $d_F$. This is especially evident on the Ni $K$--edge data. It is then a qualitative observation that in aged samples the disorder around Pd and Ni is larger in thinner PdNi layers. This is a suggestion that the degradation affects mainly the regions close to the interface since should it be a homogeneous volume effect no trend would be observed with increasing the PdNi thickness. 

The analysis of Ni XAFS data is consistent with the Pd$_x$Ni$_{1-x}$ fcc model, with a small fraction of Ni (about 10 \%) in a Ni--rich phase. The best fit parameters suggest a Pd$_x$Ni$_{1-x}$ phase with $x$ between 0.7 and 0.8 for all thicker samples (slightly different from the nominal $x=$~0.84), while it decreases around 0.65 for the thinner sample in which also a large fraction of Ni--rich phase is found.  The disorder factors roughly increases for thinner samples.
The results of the Pd $K$--edge analysis are consistent with Ni data: in particular the  Pd$_x$Ni$_{1-x}$ phase appears richer in Ni for all the samples with respect to the nominal composition and the structural disorder increases in thinner samples.
Moreover all the samples depict a fraction of Pd in the Nb--rich phase, larger in sample d2.

We must notice  that  the contributions from the Ni-- and Nb-- neighbours are often in anti--phase, this could give rise to some systematic errors on the final values. Moreover, the imposed constraints may determine a systematic inaccuracy. Thus, while absolute values for $x$ may well be affected by systematic inaccuracy, we are confident about the relative trend resulting from the EXAFS analysis as a function of $d_F$. In particular, our findings suggests that structural and compositional disorder occurs within the F layers giving rise to a deviation of the local atomic structure from the ideal one. The origin of these effects resides in the F/S interfaces, as they are larger for thinner samples. The results of  Ni and Pd EXAFS data analysis consistently suggest that the  Pd$_x$Ni$_{1-x}$ phase is systematically richer in Ni with respect to the x=0.84 nominal composition. A relatively large fraction of Pd occurs in a Nb--rich phase, suggesting Pd atoms migrate into the S layers. This finding is consistent with our previous TEM analysis,\cite{temcit} which showed the presence of heavy ions (arguably, Pd) within Nb, and in particular in the bottom Nb layer. A fraction of Ni atoms goes in a Ni--rich phase, suggesting that migration of Pd into Nb not only would leave Ni--richer Pd$_x$Ni$_{1-x}$, but can also give rise to almost pure Ni regions in the sample. 

From the EXAFS study it emerges that  substantial modifications occurred in the PdNi, F layers of aged samples, with possible effects on the Nb, S layers. We stress that the magnetic properties of the PdNi alloy change with the composition,\cite{PdNimagnetic} so that any information on the actual composition of the magnetic layer is very relevant to the understanding of the electrodynamics data

In order to clarify further the changes, we performed a detailed analysis of the compositional profile in trilayers, using the ToF--SIMS technique as described in the next Section.

\section{ToF--SIMS}
\label{TOF}

ToF--SIMS (Time--of--Flight Secondary Ion Mass Spectrometry) analysis is a very powerful technique to assess the composition of a given material\cite{Benninghoven} by controlled sputtering. The capability to detect tiny traces of chemical species within small areas (less than 0.01 mm$^2$), and with a depth resolution that approaches the nm, makes it a particularly suitable tool for the investigation of nm--thick layers, such as our SFS heterostructures, in the search for anomalies at the interfaces between layers,\cite{assantoAPL2012} or for contaminants between and inside the layers.\cite{zurloJPS2013}
\begin{figure*}[ht]
\centering
\includegraphics  [width=1.9\columnwidth] {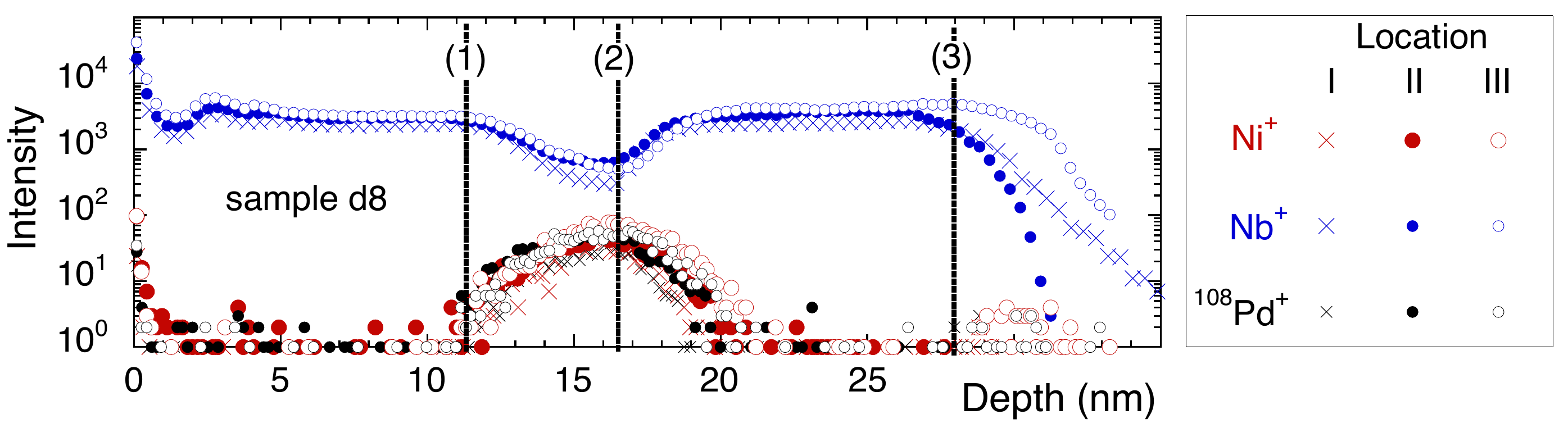}
\caption{Depth profile as measured in three different locations of the d8 trilayer. The Figure reports the Ni$^+$, Nb$^+$, and $^{108}$Pd$^+$ traces. It is seen that the behaviour is the same, an indication that the sample is homogenous along planes parallel to the layers. Vertical lines are estimates (see text) of the location of the (1) S/F, (2) F/S and (3) S/substrate interfaces. To avoid crowding, only 50\% of the data are shown.}
\label{fig:unifo} 
\end{figure*}

The  measurements were performed using a TOF.SIMS V (ION-TOF GmbH, Münster, Germany) apparatus. ToF--SIMS depth profiling have been performed in the dual-beam mode. We have analyzed samples d2, d4, d8, in order to directly compare the results to the EXAFS data and to detect systematic effects (if any) with $d_F$. The sputtering was performed using a 0.5 keV Cs$^+$
ion beam rastered over an area of 300$\times$300 $\mu$m$^2$. A 30 keV primary Bi$^+$ ion source rastered
over a scan area of 100$\times$100 $\mu$m$^2$ constitutes the analysis beam. Both ion beams were
impinging the sample surface forming an angle of 45° with the surface normal. Due to the relative complexity of the heterostructure, which contains Nb, Pd, Ni, Al (as part of the substrate, sapphire), and almost certainly Nb oxyde, we analyzed Al$^+$, Ni$^+$, Nb$^+$, NbO$^+$ and $^{108}$Pd$^+$ signals as a function of the depth in our samples. All the measurements were repeated in three different locations for each sample, to check the uniformity of the results. The sputtering time was converted to depth using a measurement of the overall thickness with a profilometer, the appearance of Nb$^+$ as zero depth and the sharp increase of Al$^+$ as a measure of the substrate level. In connection to the detection of Al$^+$, we note that, since we were also interested in evaluating the thickness of each layer, we use very little or no presputtering. Thus, close to the surface we detect all the surface contaminants, and Al among them: clearly, this is not due to the substrate. 

Before the presentation of the results, we remind that ToF--SIMS data relative to different species cannot be quantitatively compared: different intensities of the signal in different species are not necessarily a measure of the relative abundance. Moreover, primary ions, during their impact on the target surface, give rise to implantation phenomena which produce a persistency of the detection of a species even beyond its existence in the material. Thus, one should give particular care to the increase and decrease of the traces. 

We first assess the uniformity of the composition of the layers in different locations of the surface. We found very consistent results in all samples. As an illustration, in Fig. \ref{fig:unifo} we report the depth profiles for Nb$^+$, Ni$^+$ and $^{108}$Pd$^+$ in sample d8. It is seen that all traces show the same dependence on the depth: this is a demonstration that the trilayer is homogeneous over planes parallel to the surface. Moreover, it should be especially noted that Ni$^+$ and $^{108}$Pd$^+$ traces track the same behaviour.

From observation of Fig.\ref{fig:unifo}, we note that the effective thickness of the layers is consistently smaller than the nominal value. Let us focus momentarily on the Nb$^+$ traces: using the kink of the decrease as a marker for the boundary of the first (outmost) Nb layer with PdNi, we estimate $d_{Nb}\simeq$ 11 nm for the first layer. Using the increase of the same trace at a depth $\simeq$ 16 nm as the marker for the second interface, we have again $d_{Nb}\simeq$ 11 nm for the second (inner) layer, to be compared with nominal value of 15 nm. We will see in the next Figures that, as it is very reasonable, this is due to oxydized NbO layers (with O coming from the air for the outer layer, and from the substrate for the inner layer). We note that also the PdNi layer is thinner than the nominal value. Using the same criteria to evaluate the thickness, we estimate $d_{PdNi}\simeq$ 6 nm, as compared to the nominal value of 8 nm. The smooth, slow decrease of the traces above the depth of 16 nm, together with the EXAFS data, indicate that the smaller thickness of the F layer is likely to be due to diffusion processes at the lower interface PdNi/Nb.

The general features here uncovered are found in all samples. In order to discuss more thoroughly the composition of the aged structures, we now report for each sample the full set of traces (taken in a single location to avoid crowding).

\begin{figure}
\centering
\includegraphics[width=0.85\columnwidth]{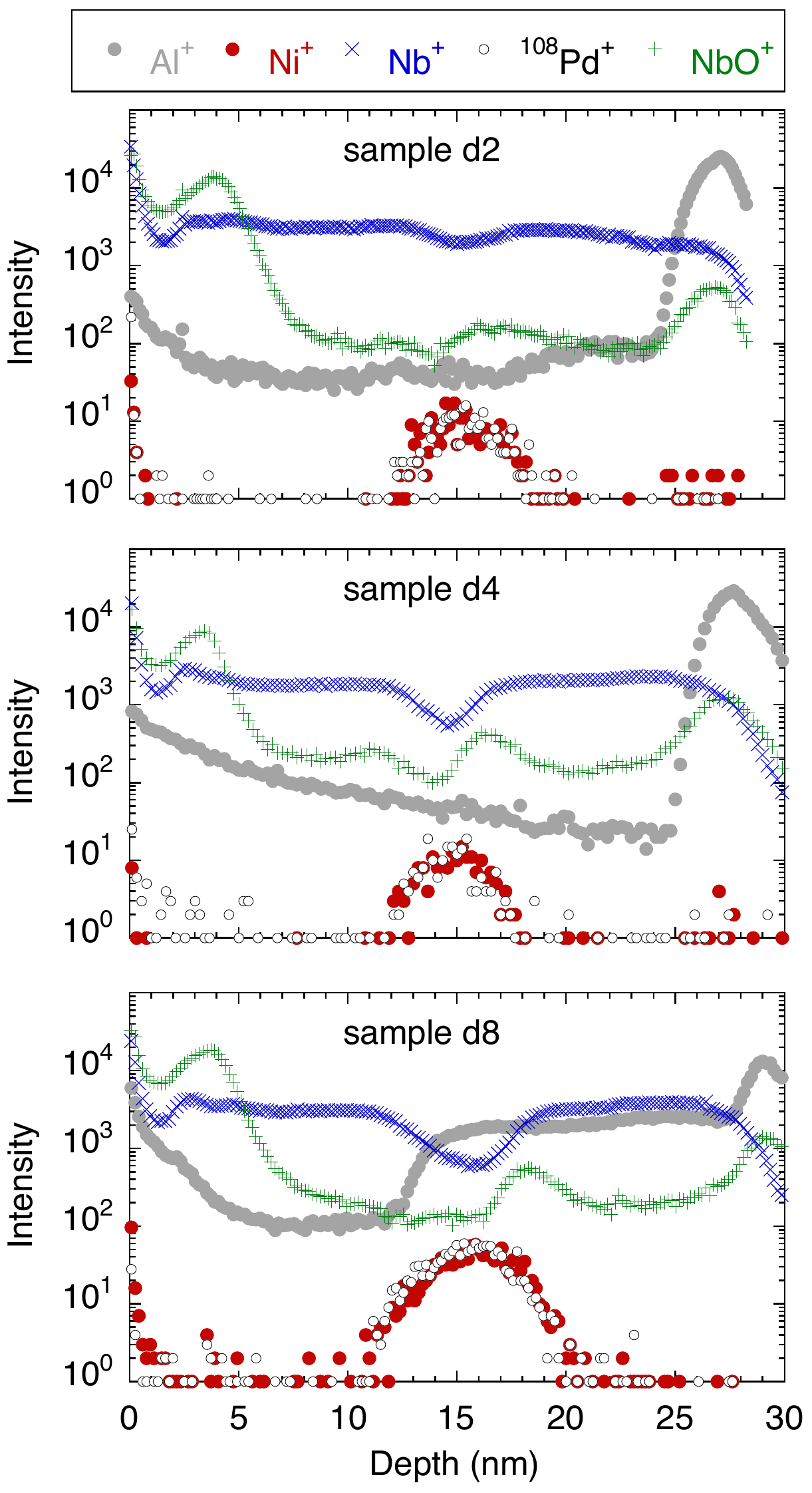}
\caption{Al$^+$, Ni$^+$, Nb$^+$, NbO$^+$ and $^{108}$Pd$^+$ signals as a function of the depth in samples d2, d4, d8 (from top to bottom). In all cases it is seen that the Nb layers have an actual thickness of $\sim$ 10 nm.}
\label{fig:ToFSIMS}
\end{figure}

Figure \ref{fig:ToFSIMS} reports in panels (a), (b) and (c) the traces for Al$^+$, Ni$^+$, Nb$^+$, NbO$^+$ and $^{108}$Pd$^+$ as a function of the depth in samples d2, d4 and d8, respectively. The Ni$^+$ and $^{108}$Pd$^+$ signals are close to the sensitivity level, but nevertheless they show the existence of the intermediate PdNi layer. The comparison with the Nb$^+$ trace allows to evaluate the PdNi thickness as $d_F\simeq$ 2, 3.5, 6 nm in samples d2, d4, d8, respectively. 

The Nb$^+$ signals decrease, but remain appreciable in the region corresponding to the PdNi layer. This is likely to be due to the collision processes, responsible for the re--implantation of ions in the sample. However, we cannot exclude that part of the signal comes from a blurred interface between Nb and PdNi. We note that the ``tails'' of the Ni$^+$ and $^{108}$Pd$^+$ signals are particularly long in sample d4. Taking into account that the measurements for Pd and Ni are very close to the sensitivity limit, we believe that this result is consistent with EXAFS data indicating diffusion of Pd into the Nb (inner) layer. 

NbO is the only significant contaminant in our samples: it is mainly present in a region which extends for $\lesssim$ 5 nm in depth, both at the air/Nb and Nb/sapphire interfaces, whence the obvious origin of Oxygen. However, we note that there is a second, smaller peak of the NbO$^+$ trace at the PdNi/Nb (inner layer) interface, most likely originating in the deposition process.

Summarizing, from ToF--SIMS measurements in aged heterostructures we have obtained that (i) the trilayers are very homogeneous over planes parallel to the surface, (ii) the actual Nb (S) thickness is $d_S\simeq$ (11 $\pm$ 1) nm, (iii) the actual PdNi (F) thickness is close, but slightly reduced with respect to nominal values, (iv) the interfaces, and in particular the F/S (inner layer) interface, is somewhat blurred over 1-2 nm. In the next and last Section we discuss all the information gathered from all the measurements up to now presented.

\section{Concluding remarks}
\label{disc}

The whole body of data collected allows for several remarks on the effect of ageing. The following discussion takes into account the effects observed on the superconducting and the $0-\pi$ transition after ageing, that is the shift to {\em higher} temperature of $T_c$ in samples d2 and d8, the shift of $T_{0\pi}$ at higher $T$ and the broadening of the transition in sample d2, and the appearance of the signature of the $0-\pi$ transition in sample d8.

We first argue that the effects of the Nb oxidisation do not play any role. In fact, the electrodynamics measurements in Nb do not show any effect at all with ageing. We note that even a simple reduction of the thickness of the superconducting layer should bring detectable changes: since the effective surface resistance in thin samples is\cite{silvaSUST96,pompeoSUST05} $R_s \simeq \Re{({1}/{\tilde{\sigma}d})}$, a reduction of the thickness $d$ would immediately bring a scale factor in the measured $R_s$, which is not seen in the data of Fig. \ref{figNb}. Thus, we conclude that the presence of NbO stems from oxidisation during deposition, for the interface Nb/sapphire, and upon exposition to the atmosphere, for the outermost layer, and that it is not the result of ageing. As a consequence, the actual Nb thicknesses of our samples have to be taken as $d_{Nb}\simeq$ 20 nm in the pure Nb film, and $d_S\simeq$ 10 nm in the trilayers.

The changes in the PdNi layer are more relevant to the $0-\pi$ transition. Let us focus first on sample d2. According to the previous findings,\cite{pompeoPRB14} the heterostructure entered the superconducting state at $T_c$ in the $0$ state, and only at lower $T_{0\pi}$ the $0-\pi$ transition occurred, to a {\em weaker} superconducting state (that is, with a smaller superfluid density). By changing the effect of the F layer (increased disorder, less clean due to mixing with the S material), the $0-\pi$ transition is broadened. The ToF--SIMS data taken in different locations of the sample stand against an explanation of the broadening in terms of local inhomogeneity. Instead, there is a clear inhomogeneity along the depth of the structure. Thus, the weakening and broadening of the $0-\pi$ transition should be ascribed first to the blurring of the interfaces (most notably, the F/S interface with the inner S layer). EXAFS also points to a change in the alloy. With a larger amount of Ni, a shorter $\xi_F$ is expected. This effect can counterbalance the deterioration of the sharpness of the interfaces and the increase of disorder, leaving a faint trace of the $0-\pi$ transition.%
%
{
We note that theoretical calculations in SFS heterostructures with a modulation of the interface predict a weakening of the sharpness of the $0-\pi$ transition, in terms of a change of the transition from first to second order.\cite{buzdinPRB2005}%
}

In sample d8 the increased disorder has a smaller effect, due to the larger F thickness. In this case, we believe that the strongest alteration is the change of the effective F thickness, from 8 nm to less than 6 nm. This effect cannot be easily mapped onto a clean system, but the appearance of the $0-\pi$ fingerprint indicates that the sample has sufficiently decreased its effective F thickness to enter the region where the superconducting state at $T_c$ is the $0$ state. The fact that $T_c$ increased significantly (by approximately 0.3 K, with $T_c\simeq$ 3.8 K in the fresh sample) also points to a smaller effect of the F layer.

In summary, the reentrance of the superconducting order parameter, fingerprint of the transition 0-$\pi$, was investigated by means of microwave measurements in  ``fresh'' and ``aged'' SFS trilayers as a function of the F layer thickness. The structural and compositional properties of the aged samples have been studied by means of EXAFS and ToF--SIMS analysis. Increased local disorder, and formation of NbPd and NiNb alloys, were found in aged samples. ToF--SIMS indicated that the interfaces were most affected. Despite severe effects in the ordering of the species in the F layers and in the quality of the SF interfaces, the temperature--induced $0-\pi$ transition is still detected. Thus, the assumption of ideal interfaces does not seem to be an essential aspect of the theoretical description,%
%
{and other physical properties play a stronger role: e.g., the exchange field and/or the ferromagnetic coherence length.\cite{kushnir}%
}%
Moreover, our data suggest that an ``effective'' F thickness can include non negligible disorder and nonideal interfaces, at least as the most relevant correction.

\section*{Acknowledgments}
LT thanks ``Fondazione Roma'' for financial support to the Surface Analysis Laboratory at Roma Tre.

\section*{Appendix A} 
The analysis of Pd and Ni $K$--edge spectra required some care because  the data statistics is generally  lower compared with Nb data, partially because of the smaller amount of absorbers that reduces the fluorescence signal. Moreover, higher structural disorder is expected due to the confinement in a thinner layer and to interface effects. In addition, the Pd and Ni atoms are randomly distributed on the fcc lattice: this raises the number of contributions to be considered, as each shell must include scattering from both Ni and Pd atoms. 

A preliminary analysis suggested a local structure around Pd and Ni different from pure Pd$_{0.84}$Ni$_{0.16}$  alloy, and  pointed to some interdiffusion effect (Nb--rich PdNb phase) and ion segregation (Ni--rich phase) that must be considered carefully to achieve a reliable quantitative interpretation of the  structural and compositional results. 
Due to the number of contributions to be considered in the analysis it is required to impose physical constraints on the fitting parameters, in order to reduce the correlation effects and to avoid unphysical results. We apply the following model:\\
{\bf{i}}. we restrict the analysis to the first two coordination shells of the PdNi phase (up to about 3.9 \AA ), each shell including both Ni and Pd neighbours;\\
{\bf{ii}}.  the total  coordination numbers of each shell were constrained to the fcc structure (N$_I$=12, N$_{II}$=6). A free parameter $x$ is defined, providing the number of Pd ($N_i^{Pd}=xN_i$) and Ni ($N_i^{Ni}=(1-x)N_i$) neighbours  for the $i$--th shell;\\
{\bf{iii}}. $N_i^{Ni}$, $N_i^{Pd}$ were further multiplied by a factor $1-Y$, $Y$ being a free parameter in the fitting representing the fraction of absorber in a  phase different from  Pd$_{x}$Ni$_{1-x}$ (see below);\\ 
{\bf{iv}}. we assume that the lattice parameter $a_{x}$ of the  Pd$_{x}$Ni$_{1-x}$ alloys has a linear behaviour as a function of the composition $x$ for $0.5<x<1$. The parameters are fixed by linear regression trough $a_{x=0.5}$, $a_{x=0.9}$ and $a_{Pd}$ values. Notice that $a_x$ is not a free parameter but constrained to $x$;\\
{\bf{v}}. for each shell the disorder factors, $\sigma^2_i$,  were kept equal for Pd and Ni neighbour.

Preliminary tests suggest that additional contributions  must be considered in the analysis of Pd and Ni XAFS, other than Pd$_{x}$Ni$_{1-x}$ phase. In this respect, Pd and Ni data behave differently:
for Pd $K$--edge data, an additional contribution is consistent with a Nb rich phase. We assume this phase having bcc structure accordingly to literature data for Nb--rich Nb$_x$Pd$_{1-x}$ phases.\cite{NbrichPd} We included the first two bcc coordination shells in the analysis, the coordination numbers were weighted by the free parameter $Y$, giving the fraction of Pd into the Nb--rich phase. 
For the Ni $K$--edge data an additional contribution is  consistent with Ni--Ni coordination shell around 2.5 \AA , consistent with fcc Ni phase ($a_{Ni}$=3.52\AA ). The fraction of Ni in the Ni--rich phase is refined using a free parameter $Y$. The disorder factor $\sigma^2_{Ni}$ was kept fixed to avoid the strong correlation with the parameter $Y$.

\end{document}